\documentclass{article}

\title{Realistic DNA De-anonymization using Phenotypic Prediction}
\date{2016-04-29}
\author{Stuart Bradley}

\usepackage{setspace}
\usepackage{multirow}
\usepackage{graphicx}
\usepackage{float}

\begin{document}
	\maketitle
	
	\begin{abstract}
		There are a number of vectors for attack when trying to link an individual to a certain DNA sequence. Phenotypic prediction is one such vector; linking DNA to an individual based on their traits. Current approaches are not overly effective, due to a number of real world considerations. This report will improve upon current phenotypic prediction, and suggest a number of methods for defending against such an attack.  
	\end{abstract}
	
	\section{Introduction}
	Human DNA is the ultimate key. Linked from a human from birth till death; it is no wonder that getting hold of such a relationship might be of paramount importance to many adversarial actors. 
	
	The reason that protection of this intrinsic information is so important, is that we are so careless about passing it around. Every place we go, every object we touch, we leave a multitude of copies. It would be an impossible task to hand this key to only those we deem worthy. Instead, striving to control the link between nucleotide and human, results in the most tractable privacy strategy. 
	
	Part of the problem with this form of defence is the value of DNA information. Altruistic parties can use such information to further improve the human condition. Hiding genetic code could result in a reduction in research speed, and a balance must be struck between saving lives, and respecting privacy. 
	
	Already, organisations like the Personal Genome Project (PGP), OpenSNP, and 23andMe curate vast databases of human genomes. OpenSNP alone has 800,000 genetic profiles \cite{Humbert-2015}. When submission occurs to any such database, participants are assured that their data is stored in an anonymised state. Practically, this entails not storing raw code with certain pieces of metadata (name and address), which in America would be in breach of the Health Information Portability and Accountability Act (HIPAA) \cite{Gymrek-2013}. What this does not account for, is that over 60\% of individuals can be identified on their ZIP code, date of birth, and sex alone \cite{Gymrek-2013}. 
	
	This article will initially look at one type of DNA de-anonymisation attack -- phenotypic prediction. This attack will be discussed in the context of real world use, and then compared with a number of other methods of DNA infiltration. These attacks will then be discussed in concert to determine the threat they pose as a joint vector, before countermeasures are discussed and evaluated. This will include the inclusion of a new countermeasure: \textit{phenotypic salt}, known colloquially, as the \textit{red-hairring}.
	
	\subsection{Briefly, biology}
	
	A general understanding of DNA, and its variation is required for a proper understanding of the topic. This section will attempt to distill human genetics into a few short paragraphs, and can be ignored by any reader well versed in the topic. 
	
	DNA -- from a computer scientist's perspective -- is a character string in a base-4 alphabet (ATCG). The string -- like binary -- is a set of encoded functions for a machine (a human) to perform higher order functions. DNA, made up of nucleotides, encodes the information to build amino acids. These in turn, link together to encode proteins. 
	
	It is at this level that we can begin to think of DNA indirectly producing macro level functions. For instance, keratin is the structural protein of hair and fingernails, amongst other things. 
	
	The reason two individuals might differ in some aspect, is due to changes in approximately 0.5\% of the genome. Physical traits (phenotypes), differ because the proteins encoding them are functionally different. Blue eyes occur due to a lack of the protein melanin \cite{Kayser-2011}, which means that the gene encoding melanin is not functioning correctly. This is likely due to a mutation in the DNA sequence itself. 
	
	It is these mutations that are often the focus in both identification and research. In the 1990s, microsatellites were used to identify persons involved in criminal cases, as well as otherwise unidentifiable bodies in disaster relief efforts and to determine familial relationships \cite{Kayser-2011}. 
	
	The specific type of microsatellite, known as a short tandem repeat (STR), is a DNA sequence between 2 and 13 nucleotides long, which repeats numerous times. The repeat number is highly variable between individuals and can therefore be used as an identifying marker.
	\begin{equation}
	\label{eq:STR}
	ATTGCATTGCATTGCATTGCATTGCATTGC
	\end{equation} 
	Equation \ref{eq:STR} is an example of a fictional STR as it is the sequence $ATTGC$ repeated six times. Part of the problem with using STRs is the amount of sequence required. In a criminal setting, DNA samples size can become restrictive to certain types of analysis. To combat this, single nucleotide polymorphisms (SNPs) can be used instead. 
	
	A SNP is a single change in a DNA sequence, which differs from individual to individual. This is one way in which we get differing physical traits. SNP mutations result in different sequences -- known as alleles -- which give rise to differing phenotypes. 
	
	\begin{equation}
	\label{eq:SNP}
	\begin{array}{c}
	ATCTTAGTCG\mathbf{G}CGATGCAG \\
	ATCTTAGTCG\mathbf{T}CGATGCAG
	\end{array}
	\end{equation}
	The example shown in Equation \ref{eq:SNP} shows that far less information is required to use SNPs for identification compared to STRs. In fact, it has been proven numerous times that significantly less DNA is required when using SNP identification methods. \cite{Kayser-2011}. 
	
	The final useful property of SNP is its relationship to the phenotype of an individual. In the case of eye colour, a single SNP (rs12913832 in the HERC2 gene), can predict an individuals eye colour, with 88\% accuracy \cite{Kayser-2011}. 
	
	\newpage
	\section{Outline of phenotypic prediction}
	In \cite{Humbert-2015}, SNPs from DNA sequences, are used to predict phenotypic traits. In essence this results in the identification of an individual from their supposedly anonymous DNA sequence. 
	
	Genotypic information can be gathered from a number of sources. Even without relying on a large scale data breach, over 800,000 individual genomes are available on OpenSNP. Phenotypic information is somewhat more difficult to access, \cite{Humbert-2015} suggests that the prevalence of open social networks allows for an adversary to gather large amounts of information, both anonymously and legally. 
	
	Using legal data sources broadens the adversarial profile dramatically. Without specific legal protections on this action, it is entirely possible that a number of legal, but ethically questionable, scenarios can occur:
	\begin{itemize}
		\item Corporations can use such an approach to link employees to their genetic profiles, furthering their ability to discriminate on a number of grounds. While this is illegal, it can be difficult to prove.
		\item Insurance companies can use a similar approach as outlined above, with respect to their customers. What makes this adversary different is in the case of health insurance, the company likely has a large amount of physical trait information that it can draw from. This removes the difficult manual step of combing through social networks to gather the necessary information. 
		\item One attack that may seem even altruistic, is the case where a researcher attempts to reach out to a person on the basis of their genetic information. Perhaps to forewarn of some predisposition to a disease. 
	\end{itemize}
	The final chunk of data is a relational database that connects SNPs to specific traits. This can be done as an unsupervised approach, using pre-existing relational databases, or as a supervised approach, which uses the collected data to build a statistical model. 
	
	Once the requisite data has been collected, \cite{Humbert-2015} outlines two possible attacks that could result in the de-anonymisation of a single person, or of a large group of people.
	
	\subsection{Identification attack}
	
	\begin{equation}
	\label{eq:ident}
	\max P(p_{x}|g_{j})
	\end{equation}
	The identification attack (Equation \ref{eq:ident}), maximizes the probability that some phenotype $p_{x}$ is caused by a genotype $g_{j}$; this results in linking the individual to a particular genotype. From a technical perspective, each genotype $g_{i}$ is ranked based on the likelihood that it is the genotype of $p_{x}$. This is done by combining the probabilities each SNP in $g_{j}$ is related to some trait in $p_{x}$. This is a computationally a very simple problem, and can be done on almost any computer. 
	
	The major downside to this attack is that it only identifies a single individual. A second proposed attack, perfect matching, attempts to deal with this problem. 
	
	\subsection{Perfect matching attack} 
	
	\begin{equation}
	\label{eq:perf_match}
	\max \prod P(p_{x}|g_{j})
	\end{equation}
	Perfect matching (Equation \ref{eq:perf_match}), attempts to maximize the match between a set of phenotypes $[P]$, to a set of genotypes $[G]$. This means that the problem increases in complexity, as every $g_{i} \in [G]$ must be ranked for every $p_{i} \in [P]$.
	
	Computationally, the problem is known as perfect matching on a weighted bipartite graph, where every phenotype is joined to every genotype by an edge. These edges are weighted based on the likelihood the genotype is related to the phenotype. The sum of these weights is then maximized to produce the most likely pairings.
	
	\subsection{Results}
	
	\begin{table}[H]
		\centering
		\caption{Percentage of matches in identification and perfect matching attacks using differing database sizes. In the unsupervised case, all data was used in the training and testing sets, as there was not sufficient data to build separate groups.}
		\label{table:1}
		\resizebox{240px}{!}{%
			\begin{tabular}{|l|llll|}
				\hline
				\multirow{2}{*}{}      & \multicolumn{2}{c|}{{\bf Database size = 80}}                                   & \multicolumn{2}{c|}{{\bf Database size = 10}}              \\ \cline{2-5} 
				& \multicolumn{1}{l|}{{\bf Unsupervised}} & \multicolumn{1}{l|}{{\bf Supervised}} & \multicolumn{1}{l|}{{\bf Unsupervised}} & {\bf Supervised} \\ \hline
				{\bf Identification}   & 5\%                                     & 13\%                                  & 44\%                                    &   52\%               \\ \cline{1-1}
				{\bf Perfect Matching} & 8\%                                     & 16\%                                  & 58\%                                    & 65\%             \\ \hline
			\end{tabular}
		}
	\end{table}
	\noindent The results of phenotypic prediction highlight its effectiveness on small datasets, such as those gathered from a single room of people. Table \ref{table:1} also eludes to the fact a statistical machine learning approach might be more effective than expertly curated databases.  
 
	\section{Criticisms of the current approach}
	
	While it is important to note \cite{Humbert-2015} was a good first step, it does not provide the scope for a complete analysis of the attack This is mostly due to the small size of databases when compared with a realistic database (800,000). This is noted in the paper, as a suggested as an improvement for further work. 
	
	Even with the small datasets, it is still possible to question two assumptions made, in an attempt to produce a more realistic picture of a phenotypic prediction.
	
	\subsection{Robustness}
	
	One problem with the attack in its current form is its accuracy. Table \ref{table:1} clearly shows that the current method in no way guarantees that an attack on an individual, or a group will be particularly successful. This is not a criticism of the attack, but of the current level of knowledge upon which the attack relies. 
	
	As knowledge of the relationships between SNPs and traits improves, so will the attack. Interestingly, the fact that the supervised approach did so well, suggests that as these attacks become more successful, they can train on the newly cracked data. Thus iteratively improving the accuracy of such an attack. However, this relies on the fact that the supervised case being better is not just an artifact of using the same data for both training and testing sets \cite{Humbert-2015}. 
	
	\subsection{Information access}
	
	When discussing the collation of information, the complexity of the task is often ignored. The inclusion of errors in any dataset, can dramatically alter the final result. 
	
	Errors in the phenotype dataset are likely to the result of human error in the gathering process. Information from social networks has no true guarantee of accuracy, and therefore could contain incorrect training information (e.g. if someone had dyed their hair). 
	
	Similarly, sequencing errors can occur in genomes. While these are painstakingly removed in the bioinformatics process, it is entirely possible they end up in the final sequencing data, and if they are in a SNP region of interest, it can be impossible to tell. The only noticeable change would be in the likelihood of the results at the end of the process.
	
	While both phenotype and genotype errors can be ignored if they are believed to be wrong, it means reducing the number of SNPs evaluated. Removing an SNP from analysis often makes it impossible to rely on the related trait; which means reducing the number of traits from 8 to 7. This is a 12.5\% reduction in the statistical power of the model, and may well result in mislabeling an individual.
	
	Errors in the SNP to phenotype database are even worse, since they result in whole groups of people being mislabeled. Since the supervised approach avoids this type of error, is more powerful.  
	
	\section{Other types of attack}
	
	Phenotype prediction is a single type of identity tracing attack. Before discussing how to combine phenotypic prediction with other forms of attack, a quick taxonomic overview should be provided \cite{Erlich-2014}.
	
	\hfill \break
	
	\subsection{Identity tracing attacks}
	
	Quasi-identifying information (such as gender, ZIP, and date of birth) can be used to identify a particular genome. As stated before, the combination of gender, ZIP and date of birth correctly identifies over 60\% of individuals. 
	
	While there are laws to protect against this form of attack (by reducing the amount of metadata that can be held in a medical database), extensive public search databases allow adversaries to link metadata to particular individuals \cite{Latanya-2013}. 
	
	Genetic genealogy is another vector for identity tracing, this is because surnames are often passed paternally, as are Y-chromosomes \cite{Gymrek-2013}. By using Y-STRs it is possible to identify 10-14\% of white American males \cite{Erlich-2014}, using publicly available genealogical databases. 	
	
	Side channel leaks attack the databases of genetic information. In the case of the PGP, files downloaded from the database containing genetic information contained full names, once unzipped \cite{Erlich-2014}. Additionally, the generation database accession numbers from identifying metadata can be attacked if the algorithm for generation is cracked.  
	
	The final type of identity tracing attack is by using pedigree charts. If familial relationships can be inferred, then any already identified genomes can be placed within this structure, reducing the search space for remaining members.	
	
	\subsection{Attribute disclosure attacks}
	
	Since DNA is reasonably prevalent in the environment, it is often entirely possible to appropriate a sample from a victim, and then use the results from a Genome Wide Association Study (GWAS), to confirm they are in control of the correct sample.
	
	To avoid this, a number of health agencies have added access control to genomes in GWAS studies. However, summary statistics remain publicly accessible.  These summary statistics are often information rich enough for the same attack to be performed \cite{Erlich-2014}.
	
	\subsection{Completion attacks}
	
	Completion attacks use already de-anonymised DNA to infer information about an anonymous genome based on \textit{genotype imputation}. This is possible because relatives may have self-identified public genomes, while the target does not. Since relatives share many markers with the target, it's entirely possible to infer large amounts of information. 
	
	This type of attack has been used successfully -- and legally -- in the Icelandic deCODE project. In this project, reference DNA was used to infer information on 200,000 individuals that had never donated their own DNA \cite{Erlich-2014}. 
	
	\section{A combinatorial approach}
	
	The statistical basis, and computational approach presented in \cite{Humbert-2015} provides a good basis for a much more complex attack. 
	
	Since a large chunk of phenotypic information is coming from social networks, it is possible to gather additional information to create a better identity tracing attack. This is discussed in the original paper, but can be taken a step further depending on the size of the database that the adversary is working from. 
	
	In the $n = 10$ case, attribute disclosure attacks (ADA) can be used. Since it's assumed a database of this size is gathered from actual samples, ADA can be employed in parallel to phenotype prediction. This helps to increase the accuracy of such an attack, by providing another method of identity tracing. Completion attacks are also beneficial in this case, since environmental samples may be significantly degraded \cite{Kayser-2011}, and inference might be necessary.
	
	In the $n = 80$ case, completion attacks aid in increasing the power of the supervised method. Initial attacks on smaller datasets can be confirmed (as in the $n = 10$ case), which increases the size of the training set for the machine learning portion of the attack. From here, additional genomes can be imputated, and the training set can increase in size. 
	
	Relationship data gathered from social media, makes possible to build pedigree charts of groups on individuals based on their familial relationships. This means that once one individual has been identified, the search space for family members becomes much smaller. This can be used in both a completion and an identity tracing approach.
	
	\section{Countermeasures}
	
	Many current suggestions for protecting against a milieu of DNA de-anonymisation attacks involve obscuring the DNA itself. Initially, it was suggested that the genomes themselves be encrypted. While this is useful for an archival approach, in many cases it increases the time to run certain analyses three-fold \cite{Erlich-2014}. Therefore, it is not acceptable for genetic research, and instead has a place in hospitals; where large scale multi-dimensional genomic research is not undertaken \cite{Humbert-2015}.
	
	Access control is another method that has been suggested \cite{Humbert-2015, Erlich-2014}. Currently, this is a widely used approach by many groups. The main problem with this, is that it slows down research due to the requirement to audit both data users, and their methods. Making sure a verified user is not doing anything nefarious with the data is still an open research area \cite{Erlich-2014}. This form of defence also arises if there is a \textit{social control} in place. If researchers decide that sharing DNA information in its current format is unethical, the solution to the problem falls outside of patchwork jurisdictions and laws, and is instead embraced by the scientific community as a larger whole. 
	
	Adding statistical noise to genomic data has also been proposed \cite{Humbert-2015}. As long as this does not impede work done on the data, it can be highly effective, and would only require a small amount of additional work to account for its presence. The attack on such a defensive measure would be akin to rainbow table on password salt. Whether an attack of this sort is feasible, depends entirely on how such noise is implemented. 
	
	The idea of statistical noise is interesting, and can be explored in the context of trait information. Where social network information is inaccurate can be seen as cryptographic salt for phenotypes. This idea can be extended to databases, so that phenotypic - or even metadata - contains salt, so that in the even of a database breach, the data is unusable (as it becomes a red-herring). 
	\begin{equation}
	\label{eq:salt}
	Haircolour: Brown + Salt \rightarrow Black
	\end{equation} 
	The problem with \textit{phenotypic salt} is that incorrect trait information (such as blood type), can wreak havoc in certain (i.e. medical) situations. Therefore, this type of defence is only applicable to the database itself. Any end user needs to see the correct (salt removed) information. 
	
	Countermeasures in genetic information need to be multifaceted in order to deal with both multiple types of attacks, as well as multiple user types. Only through the application of many measures will it be possible to keep genomic information safe. 
	
	\section{Conclusion}
	
	The initial work in \cite{Humbert-2015} proved that the de-anonymisation of DNA was a real threat. The expansion of this idea proves that it also realistic and completely possible within the near future, if not right now, in certain cases. A large amount of work still needs to be done to determine the full capabilities of such an attack. However, once achieved, they will help inform the use of countermeasures -- both technological, and legal. 
	
	\section{Acknowledgments}
	Firstly, I would like to thank Professor Clark Thomborson and Dr Rizwan Asghar for both their insight into system security, and assistance with the paper. I would also like to thank Kathleen Seddon, whose editing prowess far exceeds my own. 
	
	\bibliography{compsci725}
	\bibliographystyle{ieeetr}
\end{document}